\documentclass[pra,aps,twocolumn,epsfig,superscriptaddress,showpacs]{revtex4}
\usepackage{epsfig,amsmath}
\begin{document}
\author{Wen-Long Yang}
\affiliation{Theoretical Physics Division, Chern Institute of
Mathematics, Nankai University, Tianjin 300071, P.R.China}
\author{Jing-Ling Chen}
\email{chenjl@nankai.edu.cn} \affiliation{Theoretical Physics
Division, Chern Institute of Mathematics, Nankai University, Tianjin
300071, P.R.China}
\title {Berry's phase for coherent states of Landau levels}
\begin{abstract}
The Berry's phase for coherent states and squeezed coherent states
of Landau levels are calculated. Coherent states of Landau levels
are interpreted as a result of a magnetic flux moved adiabaticly
from infinity to a finite place on the plane. The Abelian Berry's
phase for coherent states of Landau levels is an analogue of the
Aharonov-Bohm effect. Moreover, the non-Abelian Berry's phase is
calculated for the adiabatic evolution of the magnetic field $B$.
\end{abstract}

\pacs{03.65.Vf, 03.65.-w, 47.27.De, 71.70.Di} \maketitle

\section{introduction}
Since the famous work of Berry \cite{84MVB}, the geometric phase has
been widely investigated and its generalizations have been made in
many ways \cite{83BS}\cite{84WZ}\cite{87AA}\cite{88SB}. Recently
much concern has been concentrated on the geometric phase of
entangled states as well as mixed states \cite{EP00}\cite{MP00}.
Coherent state is an important physical concept both theoretically
and experimentally \cite{g1963}\cite{creview}. It can be generated
from an arbitrary reference state, and in this Brief Report Landau
levels are chosen to be such reference states. Though coherent
states of Landau level have been studied in Refs.
\cite{LPR}\cite{SU11} and Berry's phase for coherent states as well
as squeezed coherent states of a one-dimensional harmonic oscillator
has been illustrated in Ref. \cite{87CSS}, Berry's phase for
coherent states of Landau levels which is highly degenerate and with
an additional parameter, i.e., the magnetic field $B$, is still
worthy of further investigation.

This paper is organized as follows. In Sec. II, we show how to get
the coherent states of Landau levels, and these states can be
regarded as a result of a magnetic flux moved adiabaticly from
infinity to a finite place on the plane. In Sec. III, we calculate
the Abelian and non-Abelian Berry's phase for coherent states of
Landau levels. The Abelian Berry's phase is just like an alternative
version of the Aharonov-Bohm (AB) effect; the difference between
them is that in our case the cyclic motion of the magnetic flux
results in the phase shift. In Sec. IV, we provide the explicit form
of the Hamiltonian and Berry's connections for squeezed coherent
states of Landau levels. Conclusion and discussion are made in the
last section.

\section{Coherent states of Landau levels}
The motion of a free electron in a two-dimensional $xy$-plane in a
static magnetic field along the $z$-direction is described by the
following Hamiltonian
\begin{equation}
\label{h}H=\frac{\hbar^2}{2 \mu} [ (p_x + \frac{e}{c} A_x)^2 + (p_y
+ \frac{e}{c} A_y)^2 ],
\end{equation}
where $\mu$ is the mass of the electron, $\hbar$ is the Planck
constant, $-e$ is the electron charge, $c$ is the speed of light in
the vacuum, $p_x$ and $p_y$ are the linear momentums, $A_x$ and
$A_y$ are vector potentials of the magnetic field satisfying
$\partial_x A_y -
\partial_y A_x = B$. For simplicity the Zeeman's term is not
included.

 We introduce the following operators
\begin{eqnarray}
\label{PI}\pi_x = p_x + \frac{e}{c} A_x,\; \pi_y = p_y + \frac{e}{c}
A_y,\; \pi_\pm = \pi_x \pm i \pi_y.
\end{eqnarray}
from which one can form a pair of operators $b^+$ and $b$ which
satisfy the commutation relation $[b,b^+]=1$:
\begin{eqnarray}
\label{b}b=\sqrt{\frac{c}{2 \hbar e B}}\pi_- , \;\;
b^+=\sqrt{\frac{c}{2 \hbar e B}}\pi_+ .
\end{eqnarray}
In this case we can rewrite the Hamiltonian in a more simpler form
\begin{equation}
\label{h0}H_0=\hbar \omega (b^+ b+1/2),
\end{equation}
where $\omega=e B/c\mu$, and $b^+$, $b$ are raising and lowering
operators between Landau levels with $b^+\left|n\right \rangle =
\sqrt{n+1}\left|n+1\right \rangle$, $b\left|n\right\rangle =
\sqrt{n}\left|n-1\right\rangle$. The energy of Landau levels is
$E_n=\hbar \omega (n+1/2)$. One knows that Landau levels are highly
degenerate. In the degenerate space of Landau levels we can
introduce another pair of raising and lowering operators
\begin{eqnarray}
\label{a}a&=&\sqrt{\frac{c}{2\hbar e B}}\left(-p_x - i p_y
+\frac{e}{c} A_x + i \frac{e}{c} A_y \right),\nonumber \\
\label{a^+}a^+&=&\sqrt{\frac{c}{2\hbar e B}}\left(-p_x + i p_y
+\frac{e}{c} A_x - i \frac{e}{c} A_y \right).\nonumber
\end{eqnarray}
We choose the symmetric gauge $A_x=-\frac{y}{2}B$ and
$A_y=\frac{x}{2} B$, and then $a$, $a^+$ are commutative with $b$,
$b^+$ and $[a,a^+]=1$. Therefore the Hamiltonian commutes with $a$
and $a^+$. The ground state is defined as $\left| 0,0 \right> =
\sqrt{B e/(2 \pi c \hbar )} \exp[ - B e (x^2+y^2)/(4 c \hbar)]$, and
all other eigenstates of this system can be generated from $\left|
0,0 \right>$ state with raising operators $\left| n,m \right> =
\frac{1}{\sqrt{n! m!}} b^{+^n} a^{+^m} \left| 0,0 \right>$. The
states $\left| n,m \right>$ are also orthogonal and normalized bases
for this system. States with the same $n$ are in the same energy
level, and states with the same $n$ but different $m$ stand for the
different degenerate states on the same energy level.

The coherent states of Landau levels are generated in the following
way as in \cite{87CSS}:
\begin{equation*}
\left| n( \alpha ),m \right> = \exp ( \alpha b^+ - \alpha^* b)
\left| n,m \right>,
\end{equation*}
where $\alpha= X_1 + i X_2$. The Hamiltonian for the coherent states
is:
\begin{equation}
\label{CSH}H = D(\alpha) H_0 D^+(\alpha) = \hbar \omega \left[ ( b^+
- \alpha^*) ( b - \alpha) + \frac{1}{2} \right],
\end{equation}
where $D( \alpha ) = \exp( \alpha b^+ - \alpha^* b)$, and the
eigenstates of this Hamiltonian are always combinations of the
degenerate states with the same energy:
\begin{equation}
\label{dcs}\left| n (\alpha) \right>= \sum \limits_{m} f_m \left|
n(\alpha),m \right>,
\end{equation}
where $f_m$ are arbitrary complex numbers which make $\left| n
(\alpha) \right>$ normalized. We put Eq. (\ref{b}) back to Eq.
(\ref{CSH}), this would make the Hamiltonian easier to understand,
we get
\begin{equation}
\label{CSHB}H=\frac{\hbar^2}{2 \mu} [ (\pi_x -\sqrt{\frac{2\hbar e
B}{c}}X_1)^2 + (\pi_y +\sqrt{\frac{2\hbar e B}{c}}X_2)^2 ].
\end{equation}
We found that the magnetic vector potential is added by a constant
vector potential. This can be regarded as a result of a magnetic
flux perpendicular to the plane moving adiabaticly from infinitely
far to a finitely far position on the plane. In the following  we
would like to show how we get the result.

We can assume that the added magnetic flux is a Gaussian form
magnetic field centered $( x_0 , y_0 )$, $B' = \frac{\Phi_0}{\pi
\Delta^2}
 \exp \left[-\frac{(x-x_0)^2+(y-y_0)^2}{\Delta^2}\right]$, where
 $\Delta$ is refered to as the spread or standard deviation for
the Gaussian function. And we may choose the symmetric gauge with
respect to $(x_0,y_0)$, i.e., $\left. \nabla \cdot \mathbf{A}
\right|_{x=x_0, y=y_0} = 0$, the nonsingular vector potential for
this added magnetic field is
\begin{eqnarray}
\label{AOF}
A'_x&=&\frac{\Phi_0 \{\exp[-\frac{(x-x_0)^2+(y-y_0)^2}{\Delta^2}]-1\}(y-y_0)}{2\pi[(x-x_0)^2+(y-y_0)^2]},\nonumber \\
A'_y&=&-\frac{\Phi_0
\{\exp[-\frac{(x-x_0)^2+(y-y_0)^2}{\Delta^2}]-1\}(x-x_0)}{2\pi[(x-x_0)^2+(y-y_0)^2]}.
\end{eqnarray}
One may observe that $B'$ has nothing to do with the Hamiltonian
(\ref{h}) when $x_0,y_0\rightarrow\infty$. Now we assume that the
electron in the plane is in a certain eigenstate, for example, $|
0,0 \rangle$, in this case the electron is localized near the origin
because of $\langle x \rangle=\langle 0,0 | x| 0,0 \rangle=\langle y
\rangle=0$, $\langle x^2+y^2 \rangle =2 c \hbar/Be$. Let $\Delta
> \langle \sqrt{x^2+y^2} \rangle$, so when the flux moves
adiabaticly to a place $(x_0,y_0)$ which is finitely far from the
electron (i.e. $x_0,y_0 \gg \langle\sqrt{x^2+y^2}\rangle\simeq
\sqrt{2 c \hbar/Be}$), we assume that the electron is still
distributed around the origin and $A'_x,\ A'_y$ near the origin of
the plane can be regarded as constants. Then the Hamiltonian for the
electron will be of the form (\ref{CSHB}) with
\begin{eqnarray}
\label{X1}X_1=\sqrt{\frac{e}{2\hbar B c}} \frac{\Phi_0 y_0
[\exp\left(-\frac{x_0^2+y^2_0}{\Delta^2}\right)-1] }{2\pi(x_0^2+y_0^2)},\nonumber \\
X_2=\sqrt{\frac{e}{2\hbar B c}}\frac{\Phi_0 x_0
[\exp\left(-\frac{x_0^2+y_0^2}{\Delta^2}\right)-1]}{2\pi(x_0^2+y_0^2)}.
\end{eqnarray}
This modification of the Hamiltonian also corresponds to the
following transformation $(x,y)\rightarrow (x + \delta x,y + \delta
y)$,
where
\begin{eqnarray}
\label{dx}\delta x=\frac{\Phi_0 x_0
[\exp(-\frac{x_0^2+y_0^2}{\Delta^2})-1]}{\pi B(x_0^2+y_0^2)}, \nonumber \\
\delta y=\frac{\Phi_0 y_0 [\exp(-\frac{x_0^2+y_0^2}{\Delta^2})
-1]}{\pi B(x_0^2+y_0^2)}.
\end{eqnarray}
So the state $\psi_{00}(x,y)=\left| 0,0 \right>$ will become
$\psi_{00}(x+\delta x,y+\delta y)$. Since the distribution of the
electron is near the origin, one also makes sure that $\delta x\ll
x_0,\ \delta y \ll y_0$.

The coherent states of Landau levels is nothing but the shifted
eigenstates of Landau levels in the phase space, here we assume such
a shift happens in real space $\psi_{00}(x,y) \rightarrow
\psi_{00}(x+\delta x,y+\delta y)$. When the conditions above are
satisfied, this assumption is reasonable. It is $B'$ who causes this
small shift. We may see from Eq. (\ref{dx}) that the direction of
$B'$, i.e., the sign of $\Phi_0$ is related to the direction of the
shift. When $\Phi_0>0$ the shift is parallel to the direction of the
flux, and vice versa. Interestingly, if the flux circles the
electron once, the electron will also circle the origin in a much
smaller loop once. This is the reason why the Berry's phase emerged
as we will show in the next section.

\section{ Berry's phase for coherent states of landau levels}
We know that the Landau levels are highly degenerate, and so is the
coherent states of Landau levels. Berry's phase for degenerate
states was presented in \cite{84WZ} and may have a non-Abelian
nature. We calculated the Berry's connections as follows:
\begin{eqnarray}
&&\label{me1}\left< n(\alpha),m \right| \partial_{X_1} \left|
n'(\alpha),m' \right>=( - i X_2) \delta_{n,n'} \delta_{m,m'}
\nonumber \\
&&\ \ \ +( \sqrt{n'+1}\delta_{n,n'+1} - \sqrt{n'} \delta_{n,n'-1} ) \delta_{m,m'}, \nonumber \\
&&\label{me2}\left< n(\alpha),m \right| \partial_{X_2} \left| n'
(\alpha), m' \right> = i X_1 \delta_{n,n'} \delta_{m,m'}
\nonumber \\
&&\ \ \ +( i \sqrt{n'+1} \delta_{n,n'_1} + i \sqrt{n'} \delta_{n,n'-1}) \delta_{m,m'}, \nonumber \\
&&\label{me3}\left< n(\alpha),m \right| \partial_B \left| n'
(\alpha),m' \right> = \frac{1}{2B} (\alpha \sqrt{m'}
\delta_{m,m'-1} \nonumber \\
&&\ \ \ - \alpha^* \sqrt{m'+1}\delta_{m,m'+1})\delta_{n,n'}\nonumber
\\&&\ \ + \frac{1}{2B} \sqrt{n'm'}\delta_{n,n'-1}
\delta_{m,m'-1}\nonumber
\\&&\ \ - \frac{1}{2B}\sqrt{(n'+1)(m'+1)}\delta_{n,n'+1}
\delta_{m,m'+1}.
\end{eqnarray}
In the degenerate space, i.e., between states with the same $n$, the
Berry's connections become
\begin{eqnarray}
\label{BC}&&A^{m,m'}_{X_1}= - i X_2 \delta_{m,m'},\ A^{m,m'}_{X_2}=i
X_1 \delta_{m,m'}\nonumber
\\
&&A^{m,m'}_B=\frac{1}{2B} (\alpha \sqrt{m'} \delta_{m,m'-1}-
\alpha^* \sqrt{m'+1}\delta_{m,m'+1})\nonumber.\\
\end{eqnarray}
We found that $A_{X_1}$ and $A_{X_2}$ are Abelian. With the
adiabatic theorem for degenerate states proved in \cite{AA} and to a
higher order in \cite{BB}, we know that the $|f_m|$ in Eq.
(\ref{dcs}) will not change during the arbitrary slow evolution of
$X_1$ and $X_2$. Also because the Berry's connections of $X_1$ and
$X_2$ are Abelian, $f_m$'s will gain a Berry's phase factor, which
is the same of all $m$, after an adiabatic evolution in the
$X_1$-$X_2$ plane. The Berry's phase is
\begin{eqnarray}
\label{bp}\gamma_n&=i&\oint_C\left< n(\alpha) \right| \partial_{X_1}
\left|n(\alpha) \right>dX_1 + \left< n(\alpha) \right|
\partial_{X_2} \left|n(\alpha) \right>dX_2 \nonumber\\
&=&\oint_C X_2 dX_1- X_1 dX_2=-2S,
\end{eqnarray}
where $C$ is the path of the adiabatic evolution of $(X_1,X_2)$ in
$X_1$-$X_2$ plane and $S$ is the area of $C$. This result appeared
in \cite{87CSS} for non-degenerate coherent states. The result is
also the same as the phase in the paper \cite{PE00}. However, in our
case it is the moving magnetic flux that moves the electron instead
of a moving potential well.

With the interpretation in the above section, we can see from Eq.
(\ref{X1}) that when the magnetic flux circles the electron for one
loop, the $(X_1,X_2)$ will also enclose an area, and this gives the
Berry's phase. For example, we let $(x_0,y_0)$ moves around the
origin in a circle with the radius $R$ for one loop. The Berry's
Phase will be
\begin{eqnarray}
\gamma'=-2 S'=-\frac{e \Phi_0^2(1-e^{-R^2/{\Delta}^2})^2}{4 \pi
\hbar c B R^2}.
\end{eqnarray}
This can be viewed as an alternative version of the AB effect. The
different between them is that we move the magnetic flux instead of
the electron.

One may see from Eq. (\ref{me3}) that $A^{m,m'}_B$ is non-Abelian,
so the change of $B$ will give non-Abelian Berry's phase. As the
non-Abelian Berry's phase in such a system has not been shown in the
literature before, in the following we would like to give a simple
examples to illustrate it.
\begin{figure}
\begin{center}
\epsfig{figure=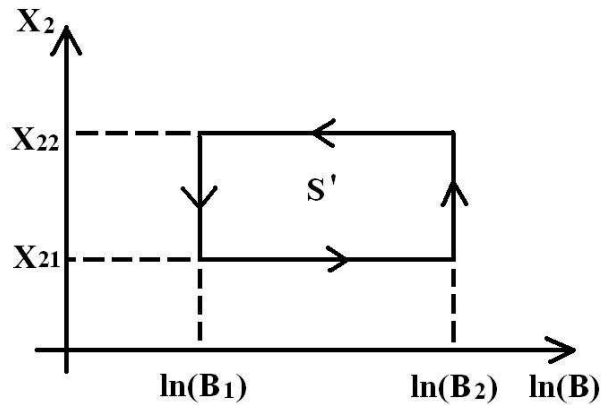,width=8cm}
\end{center}
\caption{This is the loop of the adiabatic evolution in
$X_2$-$\ln(B)$ plane. The magnetic field $B$ appears in Berry's
phase in the logarithm form.} \label{loopfig}
\end{figure}
We assume $X_1=0$ during the evolution and the other two parameters
undergo the loop in Fig. \ref{loopfig}. We can get the eigenvalues
of matrix $A_B$ as $\varepsilon_\xi/(2B)$ and the corresponding
eigenstates $\left|n(\alpha),\xi \right>$. The states before
evolution Eq. (\ref{dcs}) can be rewritten in the new base as
\begin{eqnarray}
&&\left| n(iX_{21},t=0) \right>\nonumber\\
&&=\sum \limits_{\xi} \left< n(i X_{21}),\xi \right| \left. n(i
X_{21}) \right> \left| n(i X_{21}),\xi \right>.
\end{eqnarray}
After the system undergoes an evolution as shown in Fig.
\ref{loopfig}, the state may become
\begin{eqnarray}
&&\left| n(iX_{21},t=\tau) \right>\nonumber\\
 &&=\sum \limits_{\xi}
e^{-i \varepsilon_{\xi} S' }\left< n(i X_{21}),\xi \right| \left.
n(i X_{21}) \right> \left| n(i X_{21}),\xi \right>,
\end{eqnarray}
where $S'$ is the area enclosed by $X_2$ and $\ln(B)$ as in Fig.
\ref{loopfig}. However, if $X_1\neq0$, the calculation will involve
path ordered integral and beome very complicated.

\section{Berry's phase for squeezed coherent states of Landau levels}
The Hamiltonian for squeezed coherent state is
\begin{eqnarray}
\label{SCSH}H=D(\alpha)S(\beta) H_0 S^+(\beta)D^+(\alpha),
\end{eqnarray}
where $H_0$ is defined in Eq. (\ref{h0}), $\beta = r e^{i \theta}$
and
\begin{eqnarray}
\label{s1}S(\beta)=\exp(\frac{1}{2} \beta b^{+2} -
\frac{1}{2}\beta^* b^2).
\end{eqnarray}
The eigenstates for this Hamiltonian, i.e., the squeezed coherent
states are
\begin{eqnarray}
\left| n(\alpha,\beta),m \right> = D(\alpha)S(\beta)\left| n,m
\right>.
\end{eqnarray}
In the same way,we put Eq. (\ref{b}) to Eq. (\ref{SCSH}), and we can
get
\begin{eqnarray}
\label{SCSHB}H &=&\frac{1}{2\mu}\left\{ e^{-2r} \left[\cos(\theta
/2) \pi'_x +
\sin(\theta /2)\pi'_y  \right]^2 \nonumber \right. \\
& & \left.  + e^{2r} \left[ -\sin(\theta/2)\pi'_x + \cos(\theta/2)
\pi'_y  \right]^2 \right\},
\end{eqnarray}
where $\pi'_x=\pi_x - \sqrt{2\hbar e B/c} X_1 ,\ \pi'_y=\pi_y +
\sqrt{2\hbar e B/c} X_2$. For $r=0$, Eq. (\ref{SCSHB}) reduces to
Eq. (\ref{CSHB}). For $r\neq0$, one can see
 from this Hamiltonian that the squeezing operation $S(\beta)$
caused an anisotropy in the plane. More clearly if we set $\theta=0$
Eq. (\ref{SCSHB}) will become $H=\left( e^{-2r}\pi'^2_x + e^{2r}
\pi'^2_y\right)/2\mu$, in other words, the kinetic energies
$\pi'^2_x/2\mu$, $\pi'^2_y/2\mu$ are squeezed by the factors
$e^{-2r}$ and $e^{2r}$ respectively.

Now we consider the Berry's connections of squeezed coherent states.
\begin{widetext}
\begin{eqnarray}
\label{prt}\left< n(\alpha,\beta),m \right| \frac{\partial
}{\partial r} \left| n'(\alpha,\beta),m' \right> &=&
\left[-\frac{1}{2}
 e^{-i \theta}
\sqrt{n'(n'-1)} \delta_{n,n'-2} + \frac{1}{2} e^{i \theta}\sqrt{(n'+1)(n'+2)}\delta_{n,n'+2} \right]\delta_{m,m'}, \nonumber\\
\left< n(\alpha,\beta),m \right| \frac{ \partial}{\partial \theta}
\left| n'(\alpha,\beta),m' \right> &=& \frac{i \sinh(2r)}{4} \left[
e^{-i \theta} \sqrt{n'(n'-1)}\delta_{n,n'-2} + e^{i \theta}
\sqrt{(n'+1)(n'+2)} \delta_{n,n'+2}\right]\delta_{m,m'}\nonumber
\\&+& \frac{i
\sinh^2 r}{2} (2n'+1)\delta_{n,n'}\delta_{m,m'},\nonumber\\
\label{pb}\left< n(\alpha,\beta),m \right| \frac{\partial }{
\partial B}
\left| n'(\alpha,\beta),m' \right> &=& \frac{1}{2B} \left[ 2
\sinh^2 \frac{r}{2} \sqrt{n'm'} \delta_{n,n'-1}\delta_{m,m'-1}
-2\sinh^2\frac{r}{2} \sqrt{(n'+1)(m'+1)}
\delta_{m,m'+1}\delta_{n,n'+1}
\right.\nonumber \\
&+&\left.e^{i \theta}\sinh r \sqrt{m'(n'+1)} \delta_{n,n'+1}
\delta_{m,m'-1}  - e^{-i \theta} \sinh r \sqrt{(m'+1) n'}
\delta_{m,m'+1}
\delta_{n,n'-1} \right]\nonumber \\
&+&\frac{1}{2B} (\alpha \sqrt{m'} \delta_{m,m'-1}- \alpha^*
\sqrt{m'+1}\delta_{m,m'+1})\delta_{n,n'}.
\end{eqnarray}
\end{widetext}
To our knowledge, the Berry's connection with respect to $B$ has not
been appeared in the literature before. With these, the Berry's
phase is not hard to obtain. If anisotropy exist in 2-dimensional
electron gas systems, its Hamiltonian would be of the form of Eq.
(\ref{SCSHB}).
\section{ Conclusion and Discussion}
In this Brief Report, we have calculated the Berry's phase for
coherent states as well as squeezed coherent states of Landau
levels. The Hamiltonian of the coherent states of Landau levels is
interpreted as a result of a magnetic flux perpendicular to the
plane, and it is moved adiabaticly from infinity to a distance away
from the electron so that some approximations are satisfied. The
cyclic adiabtic motion of this magnetic flux caused the Berry's
phase of coherent states of Landau levels, this is an analogue of
the AB effect. And the non-Abelian phase is also of interest, the
magnetic field $B$ appears in the Berry's phase in the form of
$\ln(B)$. So when $B\rightarrow 0$ the Berry's phase will be very
sensitive to $B$ and becomes indefinite, and the reversion of the
magnet field is prohibited if we want to get this phase. Ref.
\cite{f2005} also states this phenomenon that near the level
crossing point the Berry's phase sometimes vanishes.

We thank Prof. M. L. Ge and Prof. Jie Liu for their valuable
discussions. J. L. Chen acknowledges financial support from NSF of
China (Grant No.10605013).

\end{document}